\documentclass[12pt,a4paper]{article}
\usepackage{graphicx}
\usepackage[a4paper,left=1cm,right=1cm]{geometry}
\usepackage[latin1]{inputenc}
\usepackage{amsmath,amssymb,titling,authblk}
\usepackage{slashed}
\usepackage{amsfonts}
\usepackage{braket}
\usepackage{amsthm}
\usepackage{color}
\usepackage{dsfont}
\usepackage{slashed}
\usepackage{hyperref}
\usepackage{cite}

\newcommand{\e}{{\mathrm e}}
\newcommand{\dd}{{\mathrm d}}
\newcommand{\ii}{\mathrm i}
\newcommand{\del}{\partial}

\def\nn{\nonumber}
\newcommand{\be}{\begin{equation}}
\newcommand{\ee}{\end{equation}}
\newcommand{\bea}{\begin{eqnarray}}
\newcommand{\eea}{\end{eqnarray}}
\newcommand{\eqn}[1]{(\ref{#1})}

\newcommand{\rrho}{\lambda}

\begin{document}
\setlength{\droptitle}{-6pc}

\title{Time Discretization From Noncommutativity\vspace{5pt}}

\renewcommand\Affilfont{\itshape}
\setlength{\affilsep}{1.5em}
\renewcommand\Authands{ and }

\author[1,2,3]{Fedele Lizzi\thanks{fedele.lizzi@na.infn.it}}
\author[2,3]{Patrizia Vitale\thanks{patrizia.vitale@na.infn.it}}
\affil[1]{Dipartimento di Fisica ``Ettore Pancini'', Universit\`{a} di Napoli {\sl Federico II}\vspace{5pt}, Napoli, Italy}
\affil[2]{INFN, Sezione di Napoli, Italy\vspace{5pt}}
\affil[3]{Departament de F\'{\i}sica Qu\`antica i Astrof\'{\i}sica and Institut de C\'{\i}encies del Cosmos (ICCUB),
Universitat de Barcelona. Barcelona, Spain}

\date{}

\maketitle 

\begin{abstract}
We show that a particular noncommutative geometry, sometimes called angular or $\rho$-Minkowski,  requires that the spectrum of time be discrete. In this noncommutative space the time variable is not commuting with the angular variable in cylindrical coordinates. The possible values that the variable can take go from minus infinity to plus infinity, equally spaced by the scale of noncommmutativity. Possible self-adjoint extensions of the ``time operator'' are discussed. They give that a measurement of time can be any real value, but time intervals are still quantized.
\end{abstract}

\newpage

In general relativity spacetime itself is dynamical, therefore any theory of quantum gravity will imply  a \emph{quantum spacetime}. There are different ways to implement this quantization, and one of the most popular ones is to mimic what has been done for ordinary  quantum mechanics, and consider that the algebra generated by the coordinate functions become noncommutative, thus defining a \emph{noncommutative geometry}. Different flavours of noncommutative spaces based on noncommuting coordinates have appeared. The most promising of those have a deformed symmetry, described by a quantum group, or a Hopf algebra.

We will work in a particular kind of noncommutative spacetime in four dimensions, based on the following commutation relations among the coordinate functions:
\bea
[x^0, x^i]&=&\ii \lambda  {\epsilon^i}_{j3} x^j\nonumber \\
{[}x^i, x^j{] } &=& 0 \label{rhoMink}
\eea
where $\rrho$ is a constant with the dimensions of length and all other commutators among the coordinate functions vanish.
Notice that the third coordinate $x^3$ is central, i.e.\ it commutes with all other coordinates.
This form of noncommutativity is a particular kind of Lie algebra type noncommutativity, {the underlying Lie algebra being the Euclidean algebra,} which goes back to at least~\cite{Gutt} (also see \cite{GraciaBondia:2001ct}). In the context of twisted symmetries it was discussed by Lukierski and Woronowicz in~\cite{Lukierski:2005fc}.
In~\cite{AmelinoCamelia:2011gy} it was christened $\rho$-Minkowski because what we call $\rrho$ is called $\rho$ in that paper. We changed the notation to reserve the use of $\rho$ for the radius in cylindrical coordinates. In that paper it was shown that the principle of relative locality~\cite{AmelinoCamelia:2011bm} holds. This kind of noncommutativity might have concrete physical interest~\cite{Ciric:2019urb, Ciric:2019uab} and even phenomenological/observational consequences~\cite{Amelino-Camelia:2017pne}. A field theory of this space has been constructed in~\cite{DimitrijevicCiric:2018blz}.

Commutation relations similar to the ones considered here, and in the above references, have appeared in the definition of the ``noncommutative cylinder''~\cite{Chaichian:1998kp, Chaichian:2000ia, Dolan:2006hv, Balachandran:2007sh, Bak:2001kq, Steinacker:2011wb}. This is an example of a two-dimensional space with a compact dimension, with a commutation relation similar to~\eqref{rhoMink} (or rather~\eqref{commephi} below). In particular in~\cite{Chaichian:2000ia} time discretization, one of the results of this paper, was noted. Our discussion is however in four dimensions, there are no compact dimensions and the symmetries of the space are recovered in a quantum manner.

Emergence  of a discrete time, which is one of the main points of this paper is fascinating. Its origin goes back to no less than C.N.~Yang in 1947~\cite{Yang:1947ud}, or even earlier to Levi~\cite{Chronon} who coined the term ``Chronon".  
Discrete time also appeared in 2+1 gravity thanks to the work of 't~Hooft~\cite{tHooft:1993jwb} (see also~\cite{Balachandran:1994jq}). 
The point of view presented here is however novel, in that it connects to deformed symmetries and a promising quantum space.

We study the kinematics of this noncommutative space using the tools developed for usual quantum mechanics, namely quantise the space associating to it an algebra of  operators, obtain  a concrete representation of them on some Hilbert space, whose vectors are pure states, diagonalise sets of completely commuting observables and use the known measurement theory, namely that the possible results of a measurement are given by the eigenvalues of the observables with probabilities given by the spectral decomposition of self-adjoint operators. The word ``quantum'' in this context is ambiguous,  we use it in the sense that our space is described by operators. But Planck's action constant $\hbar$ plays no role. 
We are at a purely kinematical level. Incidentally, this will allow us to freely talk of ``time operator'', an object which in usual quantum mechanics is problematic. For a recent point of view see~\cite{Aniello:2016nvp} and references therein. If we identify $\rrho$ with Planck's length, then we are in a situation in which the inverse of the speed of light $c$ and the gravitational constant cannot be ignored, but the quantum of action can.

This kind of analysis was performed for the better known $\kappa$-Minkowski spacetime in~\cite{Lizzi:2018qaf,Lizzi:2019wto,Lizzi:2020tci}. The commutation relations in that case are of the kind
\be
[x^0,x^i]=\frac\ii\kappa x^i \ ; \ [x^i,x^j]=0
\ee
It was found that only states localised at the origin, identified with the position of a local observer, can be absolutely localised. States at a distance cannot be precisely localised. This is a consequence of an uncertainty principle which reads as
\be
\Delta x^0 \Delta x^i\geq \frac1{2\kappa} \left|\langle x^i\rangle\right|. \label{kappauncert}
\ee
Let us analyse the case of $\rho$-Minkowski. The relations~\eqref{rhoMink} are clearly of an angular nature. For this reason we work in cylindrical coordinates defined as
\be
\rho=\sqrt{{x^1}^2+{x^2}^2}\ ,\ z=\ x^3\ ,\ t=\frac{x^0}c\ ,\ \varphi=\arctan\frac{x^2}{x^1}
\ee
One could be tempted to say that in these coordinates the only nonzero commutator is
\be
[t,\varphi]=\ii\rrho
\ee
but this expression clearly does not make sense. The quantity $\varphi$ is not a single valued function and upon quantisation no self-adjoint operator would correspond to it. A correct expression is
\be
[t,\e^{\ii\varphi}]=-\ii\rrho\e^{\ii\varphi} \label{commephi}
\ee
where $\e^{\ii\varphi}$ is a legitimate well defined unitary operator.

As we said, we want to borrow the analysis from the usual quantum mechanics of point particles in three dimensions, for example. In this case we have various standard sets of mutually commuting operators. For example we can consider the three position coordinates $q^i$, and represent them as multiplicative operators on functions belonging to $L^2(\mathbb R^3)$, i.e.\ functions on configuration space. Alternatively we could consider $p_i$ as complete sets, and consider functions in Fourier transform, and position acting as a differential operator. In both these cases the spectrum is continuous (the whole line for each component), and the eigenstates are improper vectors (distributions),  Dirac $\delta$'s and plane waves respectively. Other standard choices are the three number operators 
\be
N_i=\frac12(p_i^2+q_i^2)-\frac12
\ee
In this case the spectrum is discrete and the eigenfunctions are represented by Hermite polynomials multiplied by a  Gaussian. All sort of combinations of continuous and discrete spectrum can occur; for example, for the hydrogen atom a complete set of observables is represented by the Hamiltonian itself, the square of the angular momentum  and one of its components. In this case the spectrum has continuous and discrete components, and the eigenfunctions are combinations of Laguerre polynomials, exponential functions and spherical harmonics. Any complete set will do, as long as the operators belonging to the set are self-adjoint and commute.

An important aspect to note is that the quantization of  phase space has representations on square integrable functions of a lower dimensional space. The choice of the complete set indicates which observables we may simultaneously measure. Let us consider first the case of $L^2(\mathbb R^3)$ functions and configuration space variables as a complete set of commuting observables.  
Classically the momentum of a particle is related to the velocity, proportional to it in the absence of magnetic forces.  There is no problem in the perfect localisation in position and momentum. The $\delta$'s are states of the commutative algebra of position and momentum variables. Quantum mechanically it is still possible to localise the state in position space, but to do this it is necessary to superimpose particles of all momenta: 
\be
\delta(\vec x)=\frac1{2\pi}\int_{-\infty}^\infty\!\dd^3 k\, \e^{\ii\vec k\cdot\vec x}
\ee
Good knowledge of position implies bad knowledge of momentum, and viceversa. Both position and momentum are self-adjoint non bounded operators with spectrum the real line. Up to signs and complex conjugations they are symmetrical in the theory.
Likewise, to obtain the improper eigenfunction $\delta$ as a superposition of eigenfunctions of the number operators, an infinite series is necessary, whose coefficients are not particularly simple or illuminating.  Conversely, to express the eigenfunctions of the number operator in the basis in which position is diagonal we need to give infinite information, i.e.\ a function of the variables: in this case an Hermite polynomial times a Gaussian. All this is of course well known. We just stress it to compare with the noncommutative cases below.

Let us first  review what has been done for $\kappa$-Minkowski space-time in \cite{Lizzi:2018qaf,Lizzi:2019wto,Lizzi:2020tci}. In this case  the only non trivial commutator can be expressed as 
\be
[t,r]=\frac\ii\kappa r.
\ee
A possible set of commuting coordinates  is thus given by the spatial coordinates\footnote{We have already commented on the fact that $\varphi$, an $\rrho$ are not selfadjoint operators, we nevertheless use notations like  $\{r,\rrho,\varphi\}$ as an useful shorthand. What we mean is that the observables are the  $\{x^i\}$'s acting on functions written in spherical (and later cylindrical) coordinates.}, and 
it is possible to represent $t$ as an operator acting on functions of $r$ as a \emph{dilation}:
\be
t=\frac\ii\kappa \left(r\del_r + \frac32\right)
\ee
where the $\frac32$ factor is necessary for self-adjointness.
The time operator has a continuous spectrum, and the (improper) eigenfunctions are the distributions
\be
T_\tau=\frac{r^{-\frac32-\ii\tau}}{\kappa^{\ii\tau}}=r^{-\frac32}\e^{-\ii\tau\log\left(r\kappa\right)}
\ee
which play the same role as plane waves in quantum mechanics for the operator $p$. The expansion of functions in the basis of the $\tau$ operator is therefore  provided by monomials, suggesting the use of the  \emph{Mellin} transform, which replaces the Fourier transform. Hence, a   state shall be written either as a function of $r\kappa$, or of $\tau={x^0}\kappa$, according to:
\bea
\psi(r,\theta,\varphi)&=&\frac1{\sqrt{2\pi}}\int_{-\infty}^{\infty}\dd\tau\, r^{-\frac32}\e^{-\ii\tau\log\left(r\kappa\right)} \widetilde\psi(\tau,\theta,\varphi),
\nonumber\\
\widetilde \psi(\tau,\theta,\varphi)&=&\frac1{\sqrt{2\pi}}\int_0^\infty r^2 \dd r\, r^{-\frac32}\e^{\ii\tau\log\left(r\kappa\right)} \psi(r,\theta,\varphi).
\eea
The transformation is an isometry of $L^2$, and  $|\psi|^2$ and $|\widetilde\psi|^2$ are the probability densities to find the particle at position $r$ or time $\tau$ respectively.

The uncertainty relation~\eqref{kappauncert} means that it is impossible to localise exactly a state both temporally and radially, except when the space is localised at $r=0$.
The origin is the point at which the observer is located, and is not a ``special point". Another observer will be located at its own (different) origin, and will be able to localize  states near to him.

As for space-time symmetries, let us recall that  $\kappa$-Minkowski {commutation relations are  not Poincar\'e invariant;  indeed they are } $\kappa$-Poincar\'e invariant, and translations in this case are not commuting. Therefore there is no contradiction in the fact that it is impossible for Alice to locate a state which Bob may. Alice cannot even precisely locate Bob!
To summarise, for $\kappa$-Minkowski space-time we have two complete sets of operators,
$\{x^i\}=\{r,\theta,\varphi\}$, $\{t,\theta,\varphi\}$, and the two variables $t$ and $r$  are connected by a Mellin rather than a Fourier transform. 
The noncommuting operators do not appear symmetrically. They are both unbounded with continuous spectrum, but while the spectrum of $r$ is the positive real line, that of $t$ is the whole of $\mathbb R$. The  eigenstates of the two are related by Mellin, anti-Mellin transforms, which are not symmetric as in the Fourier case. The final result is that states along the time axis can be localised, while states at a distance from the origin need to superimpose states of arbitrary time. We refer to~\cite{Lizzi:2018qaf,Lizzi:2019wto,Lizzi:2020tci} for details.

For the case of $\rho$-Minkowski there are two natural choices of complete sets of commuting observables. On the one side we have again the three position variables, which is convenient to express in cylindrical coordinates $\{\rho,z,\varphi\}$, or we may choose to have time among the observables and have $\{\rho,z,t\}$.

The three position operators act as multiplication operators, with $\rho$ and $z$ selfadjoint, and $\e^{\ii\varphi}$ unitary. The pure states of this algebra, like the previous cases, are the Dirac $\delta$ functions localised at the points of $\mathbb R^3$. It is possible to completely localise a state at any point. The time operator acts as the angular momentum in the $z$ direction. This leads to the central observation of this note:

\noindent\emph{The spectrum of time, i.e.\ the possible results of a measurement, is composed of discrete integer multiples of a quantum of time.}

\begin{figure}[htbp]\center
\includegraphics[width=0.6\textwidth]{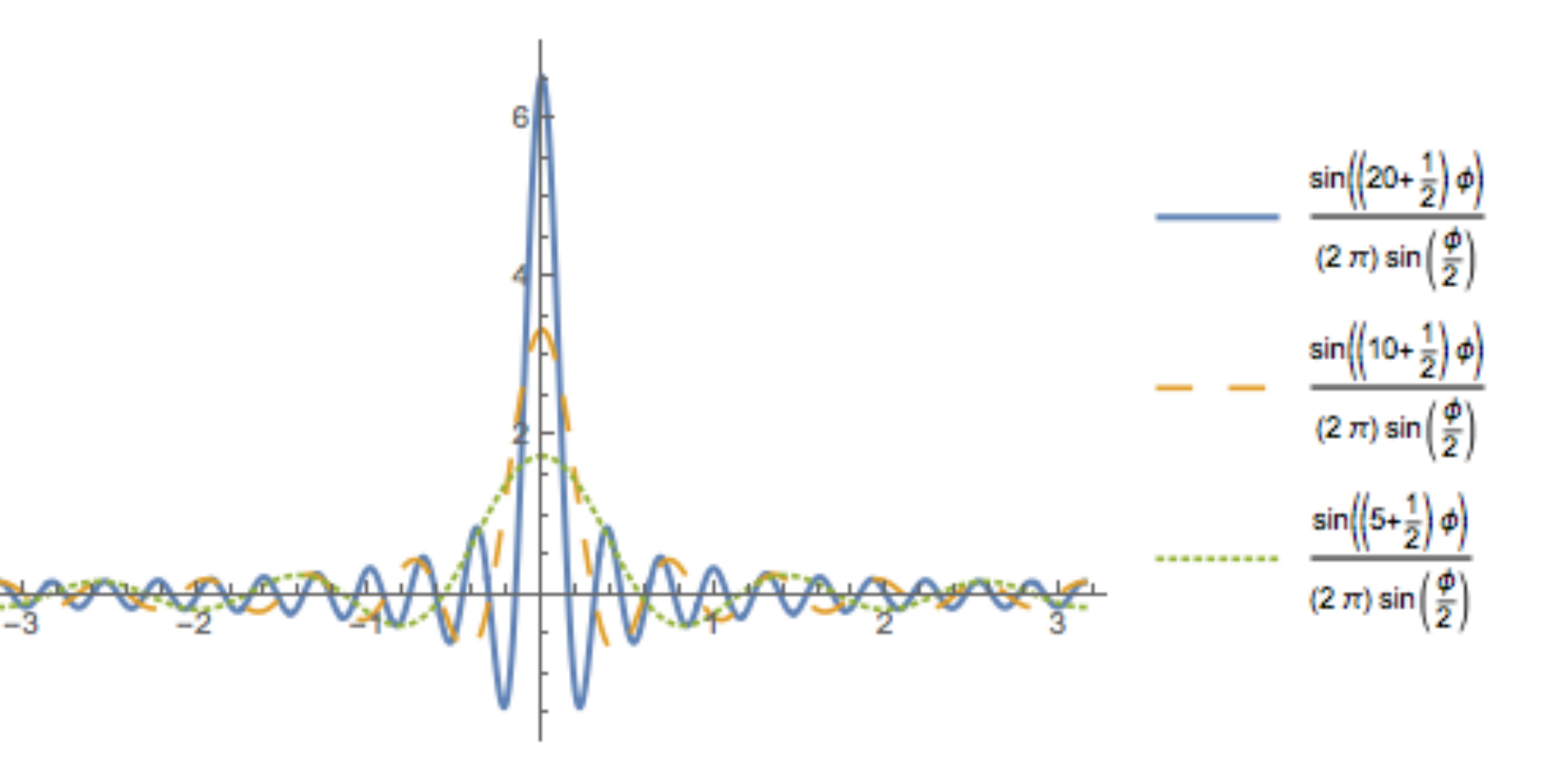}
\caption{The Dirichlet kernel for some values of $N$.} \label{Dirichl}
\end{figure}

The relation between the two bases is given by the Fourier \emph{series} expansion of the angular part:
\be
\psi(\rho,z,\varphi)=\sum_{n=-\infty}^\infty \psi_n(\rho,z) \e^{\ii n \varphi} \label{psiseries}
\ee
The time and angular variable are dual, but since $\varphi$ is not a good self-adjoint operator we cannot write the equivalent of Heisenberg uncertainty as  $\Delta t\Delta\varphi$. Nevertheless a similar reasoning can be made. As is known an eigenstate of the angular variable would need
\be
\delta(\varphi)=\frac1{2\pi}\sum_{n=-\infty}^\infty \e^{\ii n\varphi}.
\ee
On the other side, after a time measurement, which has given as result $n_0\rrho$, the system will be in an eigenstate of the time variable, namely a single $\e^{\ii n_0\varphi}$.
This means that an absolutely precise measurement of time would return a state which is uniformly dense in the angular variable.  If one measures instead time with some uncertainty, i.e.\ uses a certain number of Fourier modes to build a state peaked around some time, then the corresponding uncertainty in the angular variable is given by the fact that only a finite set of elements of the basis is available. 
If one identifies the length $\rrho$ with Planck length, converting this quantity in time units by means of the speed of light  gives for the quantum of time a quantity of the order of $5.39\ 10^{-44}\,$sec.
The most accurate measurement of time available to date is of the order of $10^{-19}\,$sec.~\cite{shorttime} Let us make a very heuristic  order of magnitude argument.  We may say that such a measurement needs the superposition of something like $10^{25}$ quanta of time. In order to localise with absolute precision the angular component of a state all infinite Fourier modes are needed, but a delta function can be well approximated by the Dirichlet nucleus
\be
\delta_N=\sum_{n=-N}^N \e^{\ii n\varphi} =\frac1{2\pi}\frac{\sin (N+\frac12) \varphi}{\sin\frac N2\varphi}.
\ee
In Fig.~\ref{Dirichl} we plot $\delta_N$ for three (extremely small!) values of $N=5,10,15$. 
This means that the most precise experiment is using $N\sim 10^{25}$. In this case the first zero of the Dirichlet nucleus (the width of the main peak of the function) is for $\varphi\sim 10^{-25}$. We may assume this to be the uncertainty in an angle determination. To translate this as an uncertainty in position we need an estimate of the radius $\rho$, the larger $\rho$, the larger the uncertainty. The  uncertainty at the edge of the observable universe ($10^{26}\,$m) is of the order of  meters. An uncertainty on the localisation of objects we can live with! 

An important aspect of $\rho$-Minkowski space-time is that,  like  $\kappa$-Minkowski~\cite{Lukierski:1991ff, Lukierski:1992dt, Majid:1994cy}, it may be regarded as the homogeneous space of a quantum group (i.e.\, a quantum Hopf algebra)~\cite{Drinfeld:1985rx}. Therefore, although its commutation relations violate standard Poincar\'e symmetry, they are covariant under the action of the appropriate deformation of the Poincar\'e group. The latter may be described following two  approaches, which are dual to each other: the Lie algebra deformation with its universal enveloping algebra and the group algebra deformation, namely the deformation of the algebra of functions on the group.  They both yield to quantum Hopf algebras, dually related, which are both referred to as the quantum group $\mathcal{G}_q$, with $\mathcal{G}$ the starting Lie group, and $q$ the deformation parameter. For the case at hand the former approach has been  described in \cite{Ciric:2017rnf},  and further analysed in \cite{DimitrijevicCiric:2018blz} in the context of field theory, whereas the latter has not been investigated up to now,  to our knowledge. Interestingly, the two points of view may be related to   observer-dependent and particle dependent transformations. Let us see how it works in the present context.

The noncommutative algebra~\eqref{rhoMink} may be realised on the algebra of functions on $\mathbb{R}^4$ in terms of a star product associated with a twist operator which reads
\bea
\mathcal{F} &=& \exp{\left\{-\frac{\ii\rrho}{2}\left[ \partial_{0}\otimes \left(x^2\partial_{1} - x^1 \partial_{2}\right)
-\left(x^2\partial_{1} - x^1 \partial_{2}\right)\otimes \partial_{0}
\right]\right\}}\nn\\
&=& \exp{\left\{\frac{\ii\rrho}{2}\left( \partial_{0}\otimes\del_{\varphi}
-\partial_{\varphi}\otimes\del_{0}
\right)\right\}} \label{angtwist0}.
\eea
The star-product is therefore defined according to
\be
(f\star g)(x):=\mu_\star (f\otimes g) (x)= \mu_0 \circ \mathcal{F}^{-1}( f\otimes g)(x).
\ee
Since the algebra is noncommutative, so is the combination of plane waves, and the sum rule of momenta:
\be
\e^{-\ii p \cdot x} \star \e^{-\ii q\cdot  x} = \e^{-\ii (R_i^j(p_j+q_j)x^i},  \label{planewaveprod}
\ee
with $R$ the following matrix:
\be
R(t) \equiv \left(
\begin{array}{cccc}
\cos{\left(\frac{\rrho t}{2}\right)} & 0 & 0 & \sin{\left(\frac{\rrho t}{2}\right)}\\
0 & 1 & 0 & 0 \\
0 & 0 & 1 & 0 \\
-\sin{\left(\frac{\rrho t}{2}\right)}  & 0 & 0 &\cos{\left(\frac{\rrho t}{2}\right)} 
\end{array}
\right) .\label{A}
\ee
In the twist approach, the Poincar\'e generators act undeformed on a single copy of the algebra of observables, with standard Lie brackets; but, in order to act on  products of observables (and therefore on the commutator~\eqn{rhoMink}), the coproduct of   Lie algebra generators  has to be twisted for the consistency of the whole enveloping algebra $\mathcal{U}(\mathfrak{p})$, according to $\Delta_{\mathcal{F}}= \mathcal{F}\Delta\mathcal{F}^{-1}$. This entails  a twisted   Leibniz rule,   so that   for $X\in \mathfrak{p}$
\be
X\triangleright f\star g:= \mu_\star\circ\Delta_{\mathcal{F}}(X)(f\otimes g)
\ee
The twisted coproduct of Poincar\'e generators~\cite{Ciric:2017rnf} is 
 given by:
\begin{eqnarray}
&&\Delta^{\cal F} P_{0} = P_{0}\otimes 1 + 1\otimes P_{0},\nn\\
&&\Delta^{\cal F} P_{3} = P_{3}\otimes 1 + 1\otimes P_{3}, \nn \\
&&\Delta^{\cal F} P_{1} = P_{1}\otimes \cos\left( \frac{\rrho}{2}P_0 \right) + \cos\left(
\frac{\rrho}{2}P_0 \right)\otimes P_{1} + P_{2}\otimes \sin\left( \frac{\rrho}{2}P_0 \right) - \sin\left(
\frac{\rrho}{2}P_0
\right)\otimes P_{2}, \nn\\
&&\Delta^{\cal F} P_{2} = P_{2}\otimes \cos\left( \frac{\rrho}{2}P_0 \right) + \cos\left(
\frac{\rrho}{2}P_0 \right)\otimes P_{2} - P_{1}\otimes \sin\left( \frac{\rrho}{2}P_0 \right) + \sin\left(
\frac{\rrho}{2}P_0
\right)\otimes P_{1}, \label{TwistedCoproductMomenta}
\end{eqnarray}
while the twisted coproduct of Lorentz generators is:
\begin{eqnarray}
&&\Delta^{\cal F} M_{31} = M_{31}\otimes \cos\left( \frac{\rrho}{2}P_0 \right) + \cos\left( \frac{\rrho}{2}P_0
\right)\otimes M_{31} + M_{32}\otimes \sin\left( \frac{\rrho}{2}P_0 \right) - \sin\left( \frac{\rrho}{2}P_0
\right)\otimes M_{32}\nn\\
&& \hspace*{1.5cm} -P_1\otimes\frac{\rrho}{2}M_{12}\cos\left( \frac{\rrho}{2}P_0 \right) +
\frac{\rrho}{2}M_{12}\cos\left( \frac{\rrho}{2}P_0 \right)\otimes P_1 \nn\\
&& \hspace*{1.5cm} -P_2\otimes\frac{\rrho}{2}M_{12}\sin\left( \frac{\rrho}{2}P_0 \right) -
\frac{\rrho}{2}M_{12}\sin\left( \frac{\rrho}{2}P_0 \right)\otimes P_2,\nn\\
&&\Delta^{\cal F} M_{32} = M_{32}\otimes \cos\left( \frac{\rrho}{2}P_0 \right) + \cos\left( \frac{\rrho}{2}P_0
\right)\otimes M_{32} - M_{31}\otimes \sin\left( \frac{\rrho}{2}P_0 \right) + \sin\left( \frac{\rrho}{2}P_0
\right)\otimes M_{31}\nn\\
&& \hspace*{1.5cm} -P_2\otimes\frac{\rrho}{2}M_{12}\cos\left( \frac{\rrho}{2}P_0 \right) +
\frac{\rrho}{2}M_{12}\cos\left( \frac{\rrho}{2}P_0 \right)\otimes P_2\nn\\
&& \hspace*{1.5cm} +P_1\otimes\frac{\rrho}{2}M_{12}\sin\left( \frac{\rrho}{2}P_0 \right) +
\frac{\rrho}{2}M_{12}\sin\left( \frac{\rrho}{2}P_0 \right)\otimes P_1,\nn\\
&&\Delta^{\cal F} M_{30}=M_{30}\otimes 1 + 1\otimes M_{30} -\frac{\rrho}{2}P_3\otimes M_{12}
+\frac{\rrho}{2}M_{12}\otimes P_3 ,\nn\\
&&\Delta^{\cal F} M_{12}=M_{12}\otimes 1+1\otimes M_{12},\nn\\
&&\Delta^{\cal F} M_{10} = M_{10}\otimes \cos\left( \frac{\rrho}{2}P_0 \right) + \cos\left( \frac{\rrho}{2}P_0
\right)\otimes M_{10} + M_{20}\otimes \sin\left( \frac{\rrho}{2}P_0 \right) - \sin\left( \frac{\rrho}{2}P_0
\right)\otimes M_{20}\nn\\
&&\Delta^{\cal F} M_{20} = M_{20}\otimes \cos\left( \frac{\rrho}{2}P_0 \right) + \cos\left( \frac{\rrho}{2}P_0
\right)\otimes M_{20} - M_{10}\otimes \sin\left( \frac{\rrho}{2}P_0 \right) + \sin\left( \frac{\rrho}{2}P_0
\right)\otimes M_{10} . \label{TwistedCoproductLorGen}
\end{eqnarray}
The 
  universal enveloping algebra of $\mathfrak{p}$, endowed with the twisted coproduct above, an antipode and a co-unit, may be shown to yield a quantum Hopf algebra, or equivalently, a quantum group, which we shall refer to as  the   $\rho$-Poincar\'e quantum group $U_\rho({\mathfrak{p}})$ (remember that the deformation parameter is $\lambda$ in our case, but we keep the notation $\rho$ to adhere to the existing literature). For the purposes of the paper we don't need to enter the technical details of the construction. What is relevant to us is that the commutation relations \eqn{rhoMink} are  twist-covariant, namely covariant under the action of the Poincar\'e generators if their action is implemented through the appropriate co-product
  \be
X\triangleright [f,  g]_*:= \mu_\star\circ\Delta_{\mathcal{F}}(X)(f\otimes g- g\otimes f).
\ee
Infinitesimal transformations of the observables, realised through the action of the deformed 
Hopf algebra $U_\lambda({\mathfrak{p}})$ are usually considered when dealing with   active transformations of physical quantities in a fixed reference frame.

Let us now consider finite transformations, which are especially relevant in connection with passive, or observer transformations \cite{Kosinski1999, Kosinski2001}. The deformation of the co-product in the Lie algebra of the group has its counterpart in the deformation of the {\it product} in the algebra of functions on the group manifold. In particular, it will affect the parameters of the Poincar\'e group, $(\Lambda^\mu_\nu, a^\mu)\in \mathcal{F}({\mathcal{P}})$.  Following~\cite{Chari} (see also~\cite{Kosinski}) we first construct  the Poisson-Lie bracket for $\mathcal{F}({\mathcal{P}})$  and then obtain the quantum group  ${\mathcal{P}}_\lambda$ by replacing the PL-bracket  with the commutator.   To this,  we  read off the $r$-matrix  from the twist operator~\eqn{angtwist0}, it being $\mathcal {F}= 1+ \frac{1}{2} r + \dots$, 
\be
r= -\ii\lambda (P_0 \wedge M_{12})  \in \mathfrak{p}\otimes\mathfrak{p}
\ee
which satisfies the classical Yang-Baxter equation \cite{Chari}. In terms of $r$ we define the Poisson-Lie bracket on the group manifold
\be
\{f , h\} = \lambda \left(X_0^R\wedge X^R_{12}-X_0^L\wedge X^L_{12}\right)(df, dg) \;\; f,g\in\mathcal{F}(\mathcal{P})
\ee
with $X_0^{R,L}, X^{R,L}_{12}$ the right and left invariant vector fields corresponding to the Lie algebra generators $P_0, M_{12}$. They read
\be
\begin{tabular}{lr}
$\begin{aligned}
 X_\alpha^L &= \Lambda^\sigma_\alpha \frac{\del}{\del a^\sigma}, \;\;\; &  X_{\alpha \beta}^L &=  \Lambda^\sigma_\alpha \frac{\del}{\del \Lambda^{\sigma\beta}}- \Lambda^\sigma_\beta \frac{\del}{\del \Lambda^{\sigma\alpha}} \\
X_\alpha^R &=  \frac{\del}{\del a^\alpha}, \; \;\;  &  X_{\alpha \beta}^R &=  \Lambda_{\beta\sigma} \frac{\del}{\del \Lambda^{\alpha}_\sigma}- \Lambda_{\alpha\sigma} \frac{\del}{\del \Lambda^{\beta}_\sigma} +a_\beta \frac{\del}{\del a_\alpha}
\end{aligned}$
\end{tabular}
\ee
and we compute 
\begin{eqnarray}
\{ {\Lambda^\mu}_\nu, {\Lambda^\rho}_\sigma \}&=&0\label{lamlam}
\\
\{ {\Lambda^\mu}_\nu, a^\rho\}&=& \lambda 
\big[-\delta^\rho_0(\Lambda_{2\nu}\delta^\mu_1
-\Lambda_{1\nu}\delta^\mu_2) 
+{\Lambda^\rho}_0( {\Lambda^\mu}_1\eta_{2\nu}- {\Lambda^\mu}_2\eta_{1\nu}) \big]\label{lama}
\\
\{a^\mu, a^\nu\}&=& \lambda \big[\delta^\mu_0(a_{2}\delta^\nu_1 -a_{1}\delta^\nu_2)   -{\delta^\nu}_0(a_{2}\delta^\mu_1 -a_{1}\delta^\mu_2) \big].
\end{eqnarray}
We thus perform the standard quantization by replacing Poisson brackets with  $-\ii$ times the commutators. We obtain
\begin{eqnarray}
[ {\Lambda^\mu}_\nu, {\Lambda^\rho}_\sigma ]&=&0
\\
{[} {\Lambda^\mu}_\nu, a^\rho {]}&=& \ii\lambda 
\big[-\delta^\rho_0(\Lambda_{2\nu}\delta^\mu_1
-\Lambda_{1\nu}\delta^\mu_2) 
+{\Lambda^\rho}_0( {\Lambda^\mu}_1\eta_{2\nu}- {\Lambda^\mu}_2\eta_{1\nu}) \big]
\\
{[}a^\mu, a^\nu{]}&=& \ii \lambda \big[\delta^\mu_0 ( a_{2}\delta^\nu_1 -a_{1}\delta^\nu_2 )   -\delta^\nu_0(a_{2}\delta^\mu_1 -a_{1}\delta^\mu_2) \big]. \label{amuanu}
\end{eqnarray}
 Since the composition law of group elements is compatible with Poisson brackets, the coproduct, antipode and counit are undeformed. The algebra $\mathcal{F}_\lambda(\mathcal{P})$ with the structures here introduced  is a quantum Hopf algebra, the quantum Poincar\'e group $\mathcal{P}_\lambda$ in the dual picture announced before. Not surprisingly, the commutator of translation parameters \eqn{amuanu} reproduces the $\rho$-Minkowski algebra \eqn{rhoMink}, the latter being identified with the  homogeneous space of the quantum Poincar\'e group with respect to the undeformed Lorentz subgroup. 
 In other words, the $\rho$-Minkowski space-time is co-acted upon according to
 \be
 {x'}^\mu= {\Lambda^\mu}_\nu\otimes x^\nu + a^\mu\otimes 1 \label{groupact}
 \ee
 and commutation relations \eqn{rhoMink} are covariant  under \eqn{groupact}, provided \eqn{lamlam}-\eqn{amuanu} hold.
 
 This completes the picture which allows for equivalent descriptions of the $\rho$-Poincar\'e group and its action on the $\rho$-Minkowski space-time. 
 
 The group algebra approach is useful to understand the observer-dependent transformations\footnote{
 A similar analysis has been performed in~\cite{Lizzi:2018qaf} for $\kappa$-Minkowski.}.
The most relevant consequence is that the transformations  relating different reference frames belong to a noncommutative algebra.   
 Hence localisability will be  subject to limitations
as well. 
States for the algebra generated by coordinates may be more or less sharply localised. When the algebra is noncommuting, there may not be states of absolute localisation. This happens  in our case.
As a consequence, different observers will not agree in
general on the localizability properties of the same state. 
We have to specify the observer making the observations, and we have been
implicitly considering an observer located at the origin. In order to change observer, a Poincar\'e transformation is needed. But in our case the symmetry is the
quantum $\rho$-Poincar\'e. Accordingly it will be impossible to locate the position of the
transformed observer, since translations do not commute.

Since time (and time slices) are discrete there appears to be a universal clock, whose beats give the allowable instants. This is  only partially true. It is known (see for example~\cite{Reed:1975uy, Esposito:2015wba}) that periodic functions are only one of the domains of selfadjointness of the operator $-\ii\del_\varphi$. A generic domain in which functions are periodic up to a phase $\e^{\ii\alpha}$ is equally good. The basis of this domain is given by functions of the kind
\be
e_\alpha=\e^{\ii(n+\alpha)\varphi}
\ee
which, together with the coefficients $\psi_n(\rho,z)$ in \eqref{psiseries}, provide an  expansion for the vectors of the Hilbert space. The spectrum of the time operator is given by the set $n+\alpha$.  The differences between eigenvalues are unchanged, and the effect is a rigid shift. Of course $\alpha$ is itself periodic of period $2\pi$. This however means that a different choice of selfadjointess domain has been made. Time translations are undeformed, and two time-translated observers will be in different, but equivalent domains. A given observer, nevertheless, can only measure quantized time intervals.

We conclude with some comments.  There are several noncommutative spaces with discrete features, reviews and references can be found for example in~\cite{Lizzi:2014pwa,DAndrea:2013rix}. Although the main motivation of this note was not to present a phenomenological viable model, nevertheless this model might be developed in a more physical direction. A field theory has been built in~\cite{DimitrijevicCiric:2018blz}, and it was shown there that decays may be affected. The $S$-matrix for these processes is discussed in~\cite{Novikov:2019kit}. In~\cite{Amelino-Camelia:2017pne} it was connected to lensing.  We feel that $\rho$-Minkowski can be added to the list of viable noncommutative spaces, and its peculiar properties deserve further investigation in a variety of directions. 

\subsubsection*{Acknowledgments}
We would like to thank Giovanni Amelino-Camelia and Jerzy Kowalski-Glikman for asking one of us to give a talk on $\rho$-Minkowski. This triggered a renewed interest in this quantum space, and inspired this work. We also wish thank Peter Schupp 
for sharing with us interesting ideas  about the interpretation of time discretisation and useful references.  We  acknowledge support from the INFN Iniziativa Specifica GeoSymQFT. FL acknowledges  the Spanish MINECO underProject No. MDM-2014-0369 of ICCUB (Unidad de Excelencia `Maria de Maeztu'), Grant No. FPA2016-76005-C2-1-P.  67985840.

\providecommand{\href}[2]{#2}\begingroup\raggedright\endgroup

\end{document}